\documentclass[10pt, conference, compsocconf]{IEEEtran}

\usepackage{amsmath}
\usepackage{algorithm}
\usepackage{algorithmic}
\usepackage{color}
\usepackage{supertabular,booktabs}
\usepackage{graphicx}
\usepackage{bm}
\usepackage[colorlinks,linkcolor=red,anchorcolor=blue,citecolor=green,CJKbookmarks=True]{hyperref}

\begin{document}

\title{A Machine Learning Framework for Stock Selection}

\author
{\IEEEauthorblockN{XingYu Fu\IEEEauthorrefmark{1}\IEEEauthorrefmark{2}\IEEEauthorrefmark{3},JinHong Du\IEEEauthorrefmark{3},YiFeng Guo\IEEEauthorrefmark{3},MingWen Liu\IEEEauthorrefmark{1}\IEEEauthorrefmark{2},Tao Dong\IEEEauthorrefmark{3},XiuWen Duan\IEEEauthorrefmark{3}}

    \IEEEauthorblockA
{
	\IEEEauthorrefmark{1}Likelihood Technology\\
}
	\IEEEauthorblockA
	{
	\IEEEauthorrefmark{2}ShiningMidas Private Fund\\
	}

\IEEEauthorblockA
{
	\IEEEauthorrefmark{3}Sun Yat-sen University\\
}
$ $\\
$\{fuxy28,dujh7,guoyf9,dongt5,duanxw3\}@mail2.sysu.edu.cn$\\
$maxwell@alphafuture.cn$

}

\maketitle

\begin{abstract}
This paper demonstrates how to apply machine learning algorithms to distinguish ``good" stocks from the ``bad" stocks. To this end, we construct 244 technical and fundamental features to characterize each stock, and label stocks according to their ranking with respect to the return-to-volatility ratio. Algorithms ranging from traditional statistical learning methods to recently popular deep learning method, e.g. Logistic Regression (LR), Random Forest (RF), Deep Neural Network (DNN), and the Stacking, are trained to solve the classification task. Genetic Algorithm (GA) is also used to implement feature selection. The effectiveness of the stock selection strategy is validated in Chinese stock market in both statistical and practical aspects, showing that: 1) Stacking outperforms other models reaching an AUC score of 0.972; 2) Genetic Algorithm picks a subset of 114 features and the prediction performances of all models remain almost unchanged after the selection procedure, which suggests some features are indeed redundant; 3) LR and DNN are radical models; RF is risk-neutral model; Stacking is somewhere between DNN and RF. 4) The portfolios constructed by our models outperform market average in back tests.
\end{abstract}


\IEEEpeerreviewmaketitle

\section{Introduction}
There are mainly three types of trading strategies to construct in financial machine learning industry, i.e. asset selection [\hyperref[ref 1]{1}], which selects potentially most profitable assets to invest, portfolio management [\hyperref[ref 2]{2}], which distributes fund into the assets to optimize the risk-profit profile, and timing [\hyperref[ref 3]{3}], which determines the apposite time to enter or leave the market. Effective asset selection is the bedrock of the whole trading system, without which the investment will be an irreparable disaster even if the most advanced portfolio management and timing strategies are deployed, and we therefore focus on asset selection, to be more specific, stock selection problem, in this article.

The essence of stock selection is to distinguish the ``good" stocks from the ``bad" stocks, which lies into the scenario of classification problem. To implement the classification system, some natural questions emerge: 1) how to label stock instances correctly? 2) what machine learning algorithms shall we choose? 3) what features to consider and how to select the best subset of features? 4) how to evaluate models' performances?

For question 1), we rank stocks according to the return-to-volatility ratio and label the top and bottom $Q$ percent stocks as positive and negative respectively [\hyperref[ref 5]{5}]. The stocks ranking in the middle of all candidates are discarded. This labeling technique enjoys two major advantages: Firstly, profitability and risk are both taken into account to give a comprehensive measure of stocks' performances, which is exactly the basic idea of Sharpe Ratio [\hyperref[ref 15]{15}]; Secondly, only the stocks whose behaviors are archetypical are used to train the classifiers, by which the ambiguous noisy information is filtered out.

For question 2), we train learning algorithms ranging from traditional statistical learning methods to recently popular deep learning method, e.g. Logistic Regression (LR) [\hyperref[ref 11]{11}], Random Forest (RF) [\hyperref[ref 10]{10}], Deep Neural Network (DNN) [\hyperref[ref 14]{14}], and the Stacking [\hyperref[ref 12]{12}], to solve the classification task. The Stacking architecture we used is a two-dimensional LR model whose inputs are the outputs of trained RF and DNN. By Stacking, we pack heterogeneous machine learning models into a single ensemble model which outperforms its every individual component. Further note that financial time series data is non-stationary and its signal-noise ratio is low, which engender the over-fitting problem [\hyperref[ref 4]{4}], especially for DNN. Hence, state-of-the-art mechanisms such as Dropout [\hyperref[ref 8]{8}], Batch-Normalization [\hyperref[ref 9]{9}] and Model Regularization [\hyperref[ref 14]{14}] are applied to DNN to alleviate the issue.

For question 3), $244$ technical and fundamental features are constructed to characterize each stock while some of the features may be redundant which introduce noise and increase the computational burden, and therefore effective features subset selection is crucial. In this work, we apply the Genetic Algorithm (GA) [\hyperref[ref 7]{7}] to conduct globally optimal subset searching in a search space with exponential cardinality, which outperforms random selection greatly in our experiments.

For question 4), the effectiveness of selection strategy is validated in Chinese stock market in both statistical and practical aspects. We first use the traditional train-test evaluation to examine the prediction performance of our models. Then we use the outputs of models to construct portfolios to trade in real historical market competing with the market average.

The evaluation result shows that: 1) Stacking of RF and DNN outperforms other models reaching an AUC score of $0.972$; 2) GA picks a subset of $114$ features and the prediction performances of all models remain almost unchanged after the selection procedure, which suggests some features are indeed redundant. 3) LR and DNN are radical investment models; RF is risk-neutral model; Stacking's risk profile is somewhere between DNN and RF. 4) The portfolios constructed by our models outperform market average in back tests.

The implementation\footnote{https://github.com/fxy96/Stock-Selection-a-Framework} of the proposed stock selection strategy is available at Github and a data sample is also shared.

\section{Dataset Construction}
\subsection{Tail and Head Label}
Say there are $M$ stock candidates and $T$ days historical stock data. We use feature vector $X_i(t) \in R^{n}$ to characterize the behavior of the $i^{th}$ stock up to the $t^{th}$ trading day and assign the return-to-volatility ratio $y_i(t,f)$ to $X_i(t)$ assessing the performance of the $i^{th}$ stock during trading period $[t+1,t+f]$, where $f$ is the length of the forward window.

For each $t$ under consideration, we obtain a set of feature vectors $\{ X_i(t) | i\in \{1,2...,M\}\}$ and a set of return-to-volatility ratios $\{ y_i(t,f) | i\in \{1,2,...,M\}\}$. We rank stocks with respect to their return-to-volatility ratios and then do the following manipulations:
\begin{itemize}
	\item Label the top $Q$ percent as positive, i.e. $y_i(t,f):=1$;
	\item Label the bottom $Q$ percent as negative, i.e. $y_i(t,f):=0$;
	\item Discard the samples in the middle.
\end{itemize}
\begin{figure}[ht]
	\centering
	\includegraphics[width=0.5\textwidth]{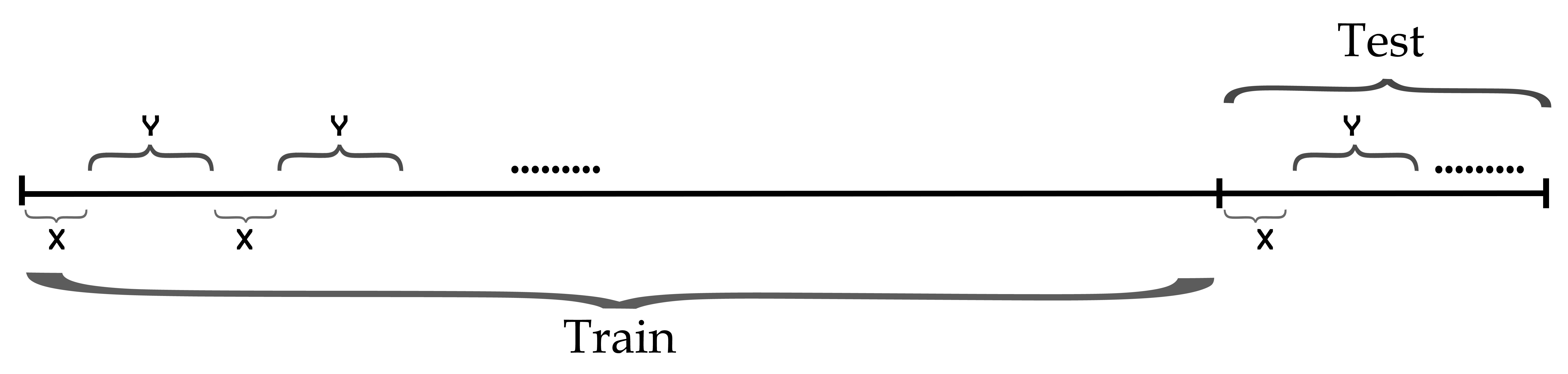}
	\caption{Data Split}\label{fig 1}
\end{figure}
To avoid data overlapping, at most $\lfloor \frac{2QTM}{100(1+f)} \rfloor$ samples can be labeled. \hyperref[fig 1]{Fig. 1.} shows how to split data.

\subsection{Feature Vector}
We construct $244$ fundamental and technical features to characterize each stock at different trading days. \hyperref[Appendix]{Appendix} lists the name of the constructed features.

\section{Genetic Algorithm based Feature Selection}
\begin{figure}[ht]
\centering
\includegraphics[scale=0.1]{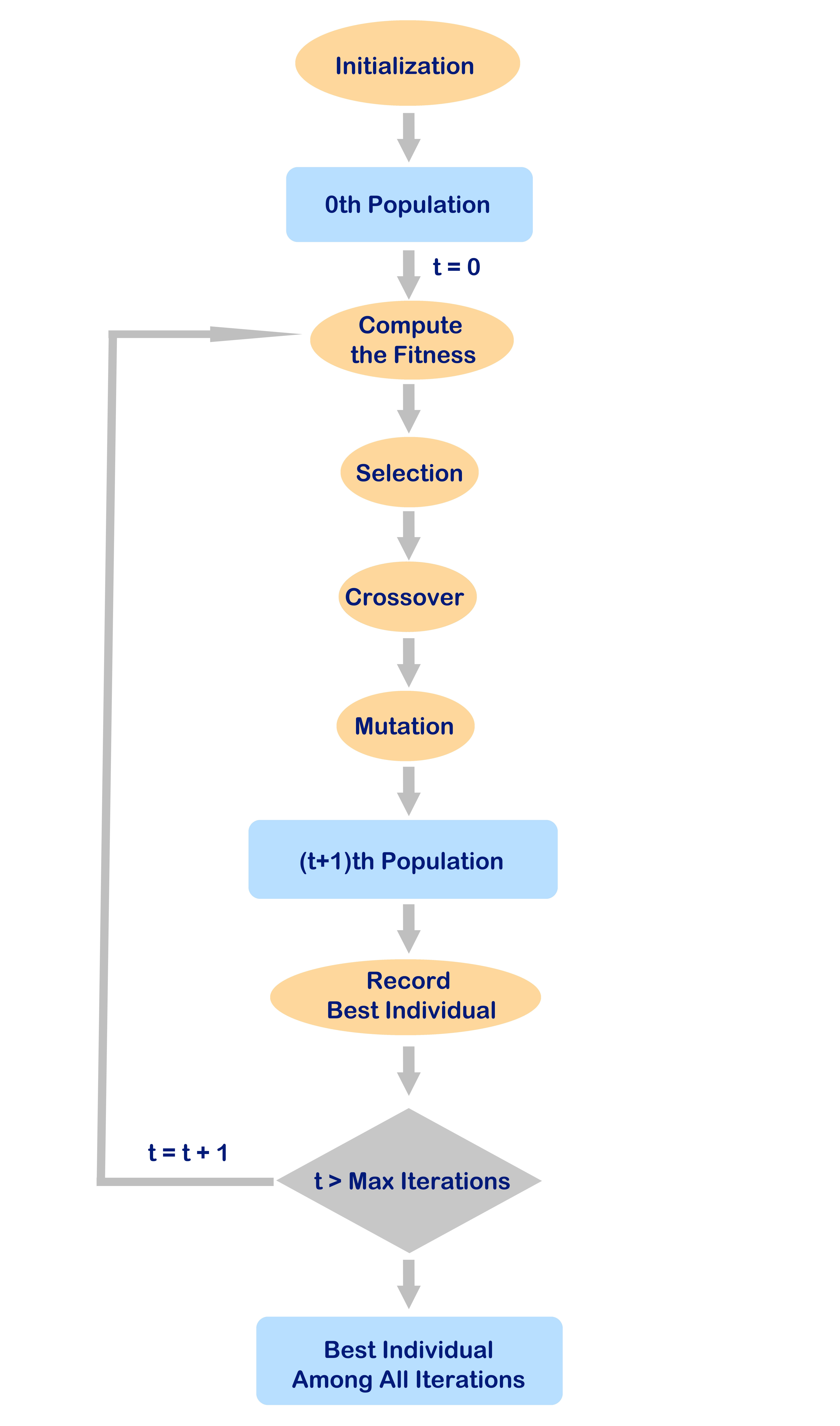}
\caption{Pipeline of GA}\label{fig 2}
\end{figure}

Genetic Algorithm (GA) is a metaheuristic inspired by Darwin Natural Selection, where a population of individuals, representing a collection of candidate solutions to an optimization problem, are evolved towards better results through iteration. Each individual is assigned with a chromosome which can be mutated or crossed. The pipeline of GA is given by \hyperref[fig 2]{Fig. 2.}

\subsection{Chromosome}
In this work, each individual's chromosome is an encoding of a subset of features. Specifically, each individual's chromosome is vector $C \in \{0,1\}^{n}$, such that $C$'s $i^{th}$ component:
\begin{equation}
C_i:=\left\{\begin{array}{rcl}
1 && \text{if the $i^{th}$ feature is selected}\\
0 && \text{else}
\end{array}\right.
\end{equation}
where $n$ denotes the total number of features, which is $244$ in our case.

\subsection{Initialization}
Suppose that the population size is $100$.
We initialize the $0^{th}$ generation by randomly generating $100$ binary vectors independently identically distributed with $Binomial(244,0.5)$.

For each individual, we need to check if the sum of its chromosome's components is not less than $1$ to ensure the subset it represents is not empty. If the condition doesn't hold, we need to regenerate this individual.

\subsection{Fitness}
To assess how good an individual is, we need to assign fitness to each individual. An individual is defined to be superior to the other if its fitness is bigger.

In our work, we define the fitness of an individual as the AUC score of a machine learning model trained on the corresponding features subset.

To be specific, for each individual, we use its chromosome to obtain a features subset and we train a logistic regression model on the training dataset only considering the corresponding features subset. Then we evaluate the LR model on the testing set and get the AUC value, which is set to be the fitness of the individual.

\subsection{Selection}
After the computation of finesses, we can construct a probability distribution over the population by normalizing all individuals' fitnesses and we select elites with respect to the distribution. Specifically, we randomly sample $100$ individuals from the current population with replacement to form the next generation. The individuals with bigger fitnesses have bigger opportunities to be picked out.

\subsection{Crossover}
For each pair of sampled individuals obtained from the previous selection procedure, with probability $0.2$ we implement the crossover mechanism, i.e. we randomly select a position on chromosome and exchange the genes of the two individuals at this position.

\subsection{Mutation}
After crossover, for each sampled individuals, with probability $0.1$ we carry out the mutation procedure, i.e. we randomly select a component on this individual's chromosome and alter its value.

\subsection{Evolution}
In each iteration, we repeat the above manipulations and record the best individual. Generally, the average fitness of the population will become increasing better as the iteration goes on and this is the reason why we call GA as evolutionary process. After $100$ iterations, we terminate the evolution and selects the best individual as the best features subset representation.

\section{Predictive Models}
In this section, we will discuss the details of models we use in our work, which are: Logistic Regression (LR), Random Forest (RF), Deep Neural Network (DNN), and the Stacking of RF and DNN.

\subsection{Logistic Regression}
LR is a classical statistical learning algorithm, which models the conditional probability of positive sample given the feature vector using the logistic function, i.e.
\begin{equation}
\bm{P}\{ Y=1 | X=x \} = \sigma(\beta\cdot x+b)
\end{equation}
, where $\sigma(t)=\frac{1}{1+e^{-t}}$ is the logistic function and $\beta \in R^n$, $b \in R$ are the linear coefficients.

The key to implement LR is to estimate the linear coefficients from historical labeled data, where in this paper, we apply the Stochastic Gradient Descent (SGD) with Nesterov momentum and learning rate decay [\hyperref[ref 16]{16}] algorithm to minimize the log-likelihood loss function to this purpose.

Although LR is a rather simple model, which results in relatively poor predictive ability, yet it is extensively used in financial machine learning industry since it is unlikely to suffer from over-fitting. \hyperref[Table 1]{Table 1} shows the hyperparameters of LR.
\begin{table}[!ht]
	\centering
	\begin{tabular}{ll}
		\hline
		\textbf{Hyperparameter} & \textbf{Value} \\
		\hline
		Initial Learning Rate  & $10^{-2}$\\
		Learning Rate Decay Rate &$10^{-6}$\\
		Training Epochs &$20$\\
		Momentum Coefficient & $0.9$\\
		\hline\\
	\end{tabular}
	\caption{LR Hyperparameters}\label{Table 1}
\end{table}

\subsection{Random Forest}
RF is a state-of-the-art machine learning technique that trains a collection of decision trees and makes classification prediction by averaging the output of each tree. The trees in forest are trained on different bootstrapping samples of training dataset and focus on different random feature subsets to break model correlation.

It is proven, by both industry and academia, that RF can effectively decrease variance of model without increasing bias, i.e. conquers over-fitting problem, and therefore RF is a promising predictive method in finance machine learning industry, where over-fitting is a common problem. \hyperref[Table 2]{Table 2} shows the hyperparameters of RF.
\begin{table}[!ht]
	\centering
	\begin{tabular}{ll}
		\hline
		\textbf{Hyperparameter} & \textbf{Value} \\
		\hline
		Number of Trees  & $100$\\
		Maximal Tree Depth &$4$\\
		Minimal Samples Number of Split &$2$\\
		Minimal Impurity of Split &$10^{-7}$\\
		\hline\\
	\end{tabular}
	\caption{RF Hyperparameters}\label{Table 2}
\end{table}
\subsection{Deep Neural Network}
DNN, a technique that recently witnessed tremendous success in various tasks such as computer vision, speech recognition and gaming [\hyperref[ref 17]{17},\hyperref[ref 18]{18},\hyperref[ref 19]{19}], is a potential power predictor in financial application. While DNN is prone to be stuck in poor local optimum if the training environment is highly noisy, and therefore effective mechanisms preventing over-fitting must be deployed if we want to use DNN in financial market.

In this work, we implement a three-layer neural network to classify stocks, where Dropout, Batch-Normalization and $L_2$ Regularization are used to avoid over-fitting. Its training is done through SGD with Nesterov momentum and learning rate decay. For most cases, the network converges after $20$ epochs. \hyperref[Table 3]{Table 3} and \hyperref[Table 4]{Table 4} demonstrate network architecture and hyperparameters respectively.
\begin{table}[!ht]
	\centering
	\begin{tabular}{ll}
		\hline
		\textbf{Layer} & \textbf{Shape} \\
		\hline
		Input Tensor & $128\times n$\\
		Fully Connected Layer with $L_2$ Regularizer & $128\times \lfloor \frac{n}{2}\rfloor$\\
		ReLu Activation & $128\times \lfloor \frac{n}{2}\rfloor$\\
		Dropout Layer  & $128\times \lfloor \frac{n}{2}\rfloor$\\
		Fully Connected Layer with $L_2$ Regularizer & $128\times \lfloor \frac{n}{4}\rfloor$\\
		ReLu Activation & $128\times \lfloor \frac{n}{4}\rfloor$\\
		Batch-Normalization Layer & $128\times \lfloor \frac{n}{4}\rfloor$\\
		Fully Connected Layer with $L_2$ Regularizer & $128\times 2$\\
		Softmax Activation & $128\times 2$ \\
		\hline\\
	\end{tabular}
	\caption{Network Struction}\label{Table 3}
\end{table}
\begin{table}[!ht]
	\centering
	\begin{tabular}{ll}
		\hline
		\textbf{Hyperparameter} & \textbf{Value} \\
		\hline
		Dropout Rate & $0.5$\\
		$L_2$ Penalty Coefficient & $0.01$\\
		Momentum Coefficient & $0.9$\\
		Initial Learning Rate  & $10^{-3}$\\
		Learning Rate Decay Rate &$10^{-6}$\\
		Training Epochs &$20$\\
		\hline\\
	\end{tabular}
	\caption{Network Hyperparameters}\label{Table 4}
\end{table}
\subsection{Stacking of RF and DNN}
Stacking is a technique to ensemble multiple learning algorithms, where a meta-level algorithm is trained to make a final prediction using the outputs of based-level algorithms as features. Generally, the stacking model will outperform its each based-level model due to its smoothing nature and ability to credit the classifiers which perform well and discredit those which predict badly. Theoretically, stacking is most effective if its based-level algorithms are uncorrelated.

In this work, RF and DNN are used as based-level classifiers and LR is selected as meta-level algorithm. We first partition the original training dataset into three disjoint sets, named as Train-1, Train-2, and Validation.

Train-1 and Train-2 serve as the training set of based-level classifiers, in which the DNN is trained on the concatenation of Train-1 and Train-2 and RF is trained solely on Train-2. By training like this, the two based-level algorithms' behaviors will be significantly uncorrelated , which is the prerequisite for a well-performed stacking, for two reasons: 1) RF and DNN belong to, in nature, two utterly different types of algorithms; 2) DNN is trained in a "Fore-Sighted" way while RF is trained in a "Short-Sighted" way. Validation is then used to train the meta-level LR model, where the inputs to LR is the predictions from trained-DNN and trained-RF on the Validation.

\begin{figure}[ht]
 	\centering
 	\includegraphics[scale=0.075]{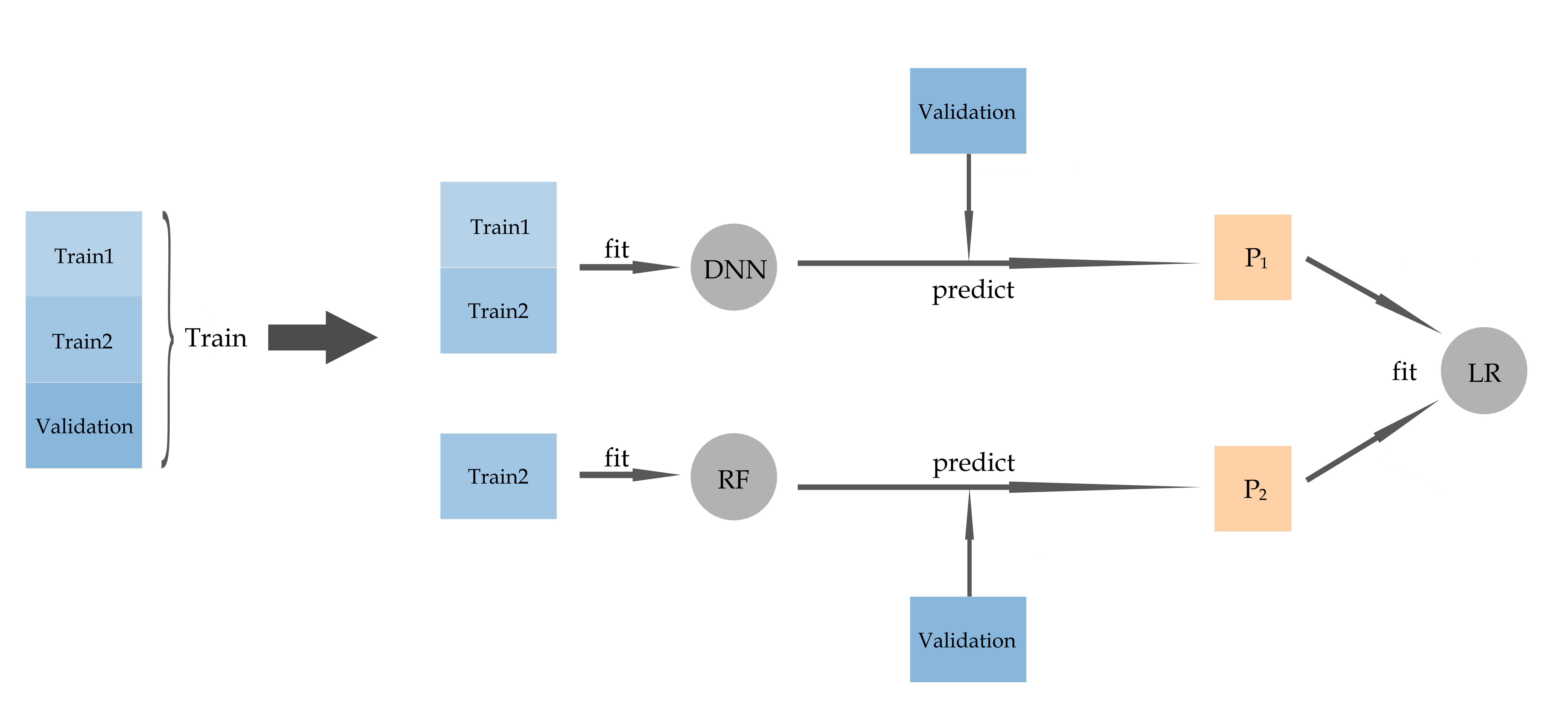}
 	\caption{Pipeline of Training Stacking}\label{fig 3}
 \end{figure}

 The pipeline of training our stacking model is shown in \hyperref[fig 3]{Fig. 3.}.

\section{Result Presentation and Empirical Analysis}
In this section, we present our result and analyze it from both statistical aspect and practical (portfolio management) aspect.

\subsection{Statistical Analysis}
\begin{figure}[ht]
	\centering
	\includegraphics[scale=0.07]{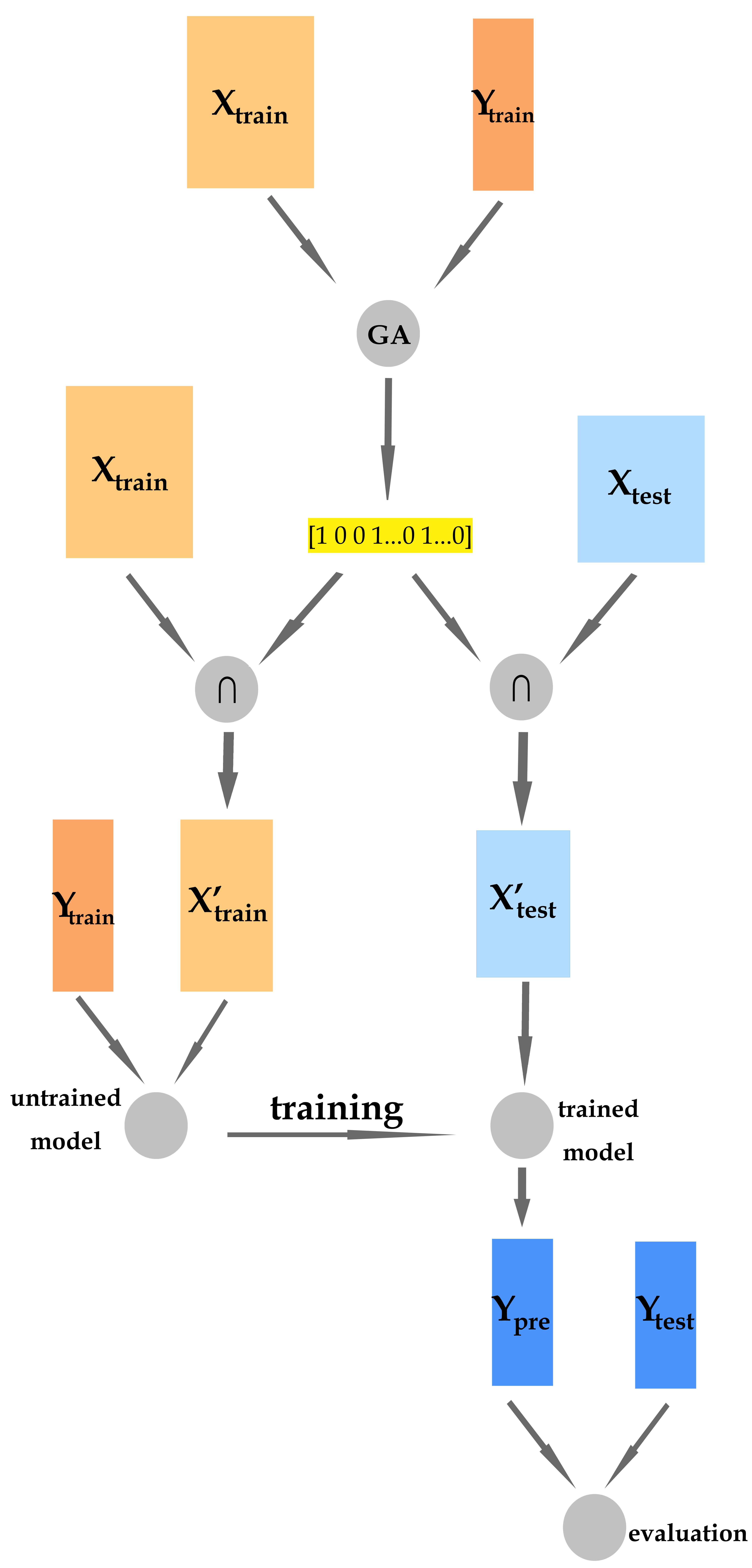}
	\caption{Pipeline of Statistical Evaluation}\label{fig 4}
\end{figure}

\hyperref[Table 5]{Table 5} and \hyperref[Table 6]{Table 6} demonstrate the statistical indexes of different models before and after GA feature selection. The training dataset is constructed from Chinese stock data ranging from 2012.08.08 to 2013.02.01 and the testing dataset is constructed from 2013.02.04 to 2013.03.08. \hyperref[fig 4]{Fig. 4.} shows the pipeline of our statistical evaluation procedure.
\begin{table}[!htbp]
	\centering
	\begin{tabular}{ccccc}
		\hline
		\textbf{} &LR&RF&DNN&Stack\\
		\hline
		AUC&0.964&0.965&0.970&0.972\\
		Accuracy&0.903&0.918&0.913&0.919\\
		Precision&0.871&0.916&0.875&0.904\\
		Recall&0.945&0.922&0.963&0.938\\
		F1&0.907&0.919&0.917&0.921\\
		TPR&0.945&0.922&0.963&0.938\\
		FPR&0.139&0.084&0.138&0.099\\
		\hline\\
	\end{tabular}
	\caption{Statistical Indexes before Feature Selection}\label{Table 5}
\end{table}
\begin{table}[!htbp]
	\centering
	\begin{tabular}{ccccc}
		\hline
		\textbf{}&LR&RF&DNN&Stack\\
		\hline
		AUC&0.966&0.970&0.958&0.973\\
		Accuracy&0.908&0.918&0.869&0.927\\
		Precision&0.886&0.919&0.819&0.919\\
		Recall&0.936&0.916&0.946&0.936\\
		F1&0.911&0.918&0.878&0.928\\
		TPR&0.936&0.916&0.946&0.936\\
		FPR&0.119&0.079&0.207&0.081\\
		\hline\\
	\end{tabular}
\caption{Statistical Indexes after Feature Selection}\label{Table 6}
\end{table}

As we can see from \hyperref[Table 5]{Table 5} and \hyperref[Table 6]{Table 6}:
\begin{itemize}
	\item Stacking of DNN and RF outperforms other models before and after feature selection reaching the highest AUC score, which shows the promising prospect of ensemble learning in financial market.
	\item All the statistical indexes remain almost unchanged before and after feature selection, which shows some features are indeed redundant.
	\item For both LR and DNN models, their Recall scores are notably higher than their Precision scores, which indicates, in plain English, LR and DNN are radical investment models. While for RF, its Recall score is commensurate to its Precision score, suggesting that RF is more likely to be a risk-neutral investment model. As for Stacking, it's risk profile is somewhere between RF and DNN since it is a combination of RF and DNN.
\end{itemize}

\subsection{Portfolio Analysis}
There is a setback of our statistical evaluation procedure, i.e. only the stocks in the $Q$ percent tail and head are included in the test sets, which means we only test our model in the unequivocal test samples. Therefore, a more close-to-reality evaluation is needed.

\begin{figure}[!ht]
	\centering
	\includegraphics[width=0.5\textwidth]{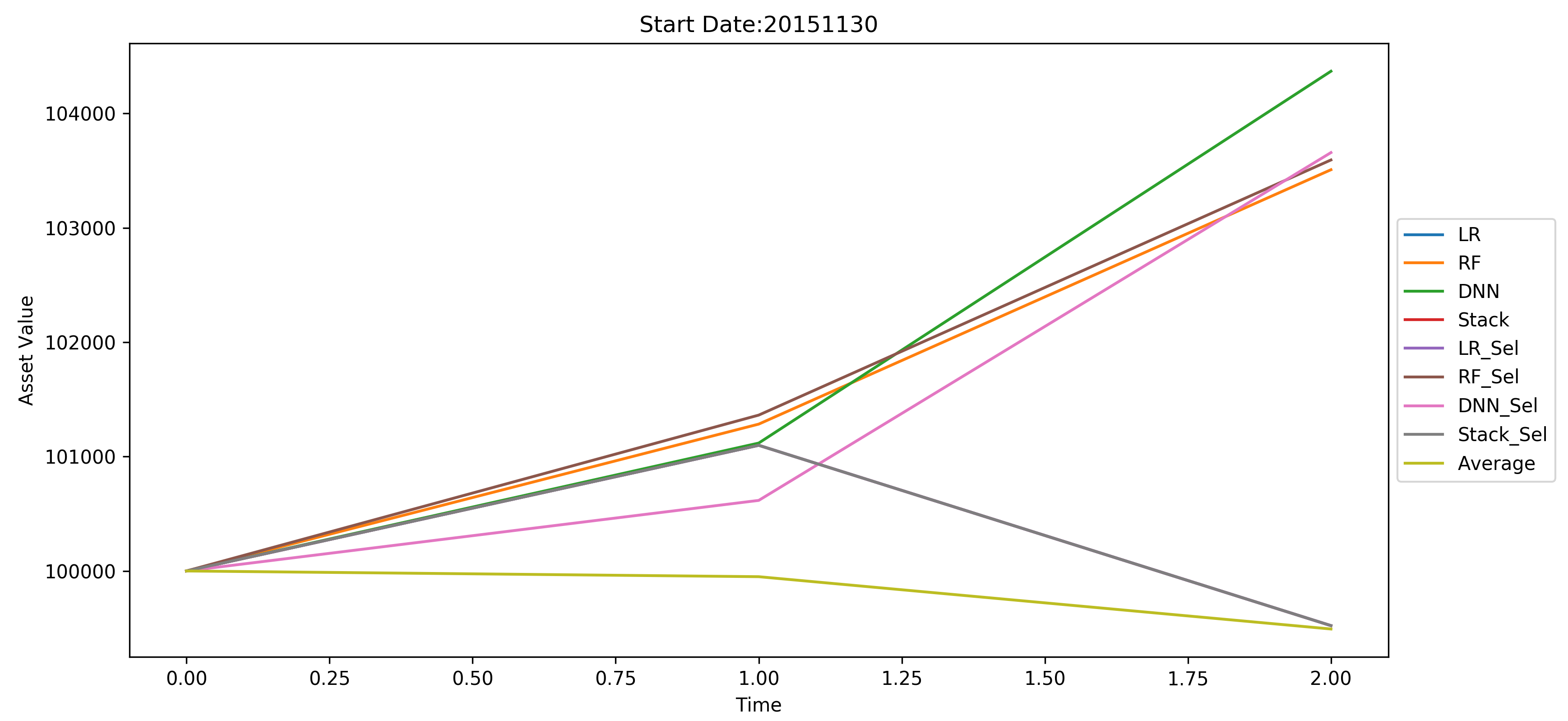}
	\caption{Portfolio Management starts from 2015.11.30}\label{fig 5}
\end{figure}

\begin{figure}[!ht]
	\centering
	\includegraphics[width=0.5\textwidth]{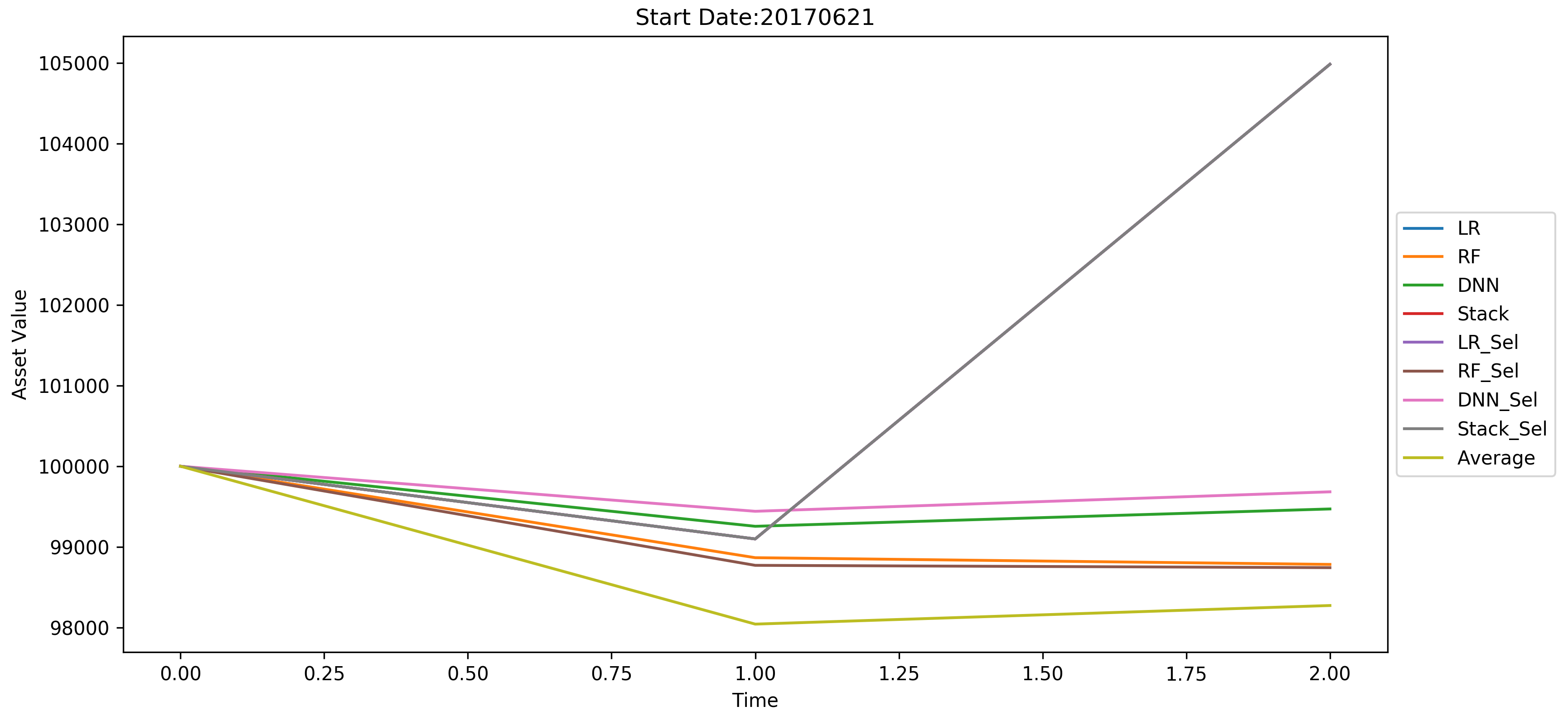}
	\caption{Portfolio Management starts from 2017.06.21}\label{fig 6}
\end{figure}

\begin{figure}[!ht]
	\centering
	\includegraphics[width=0.5\textwidth]{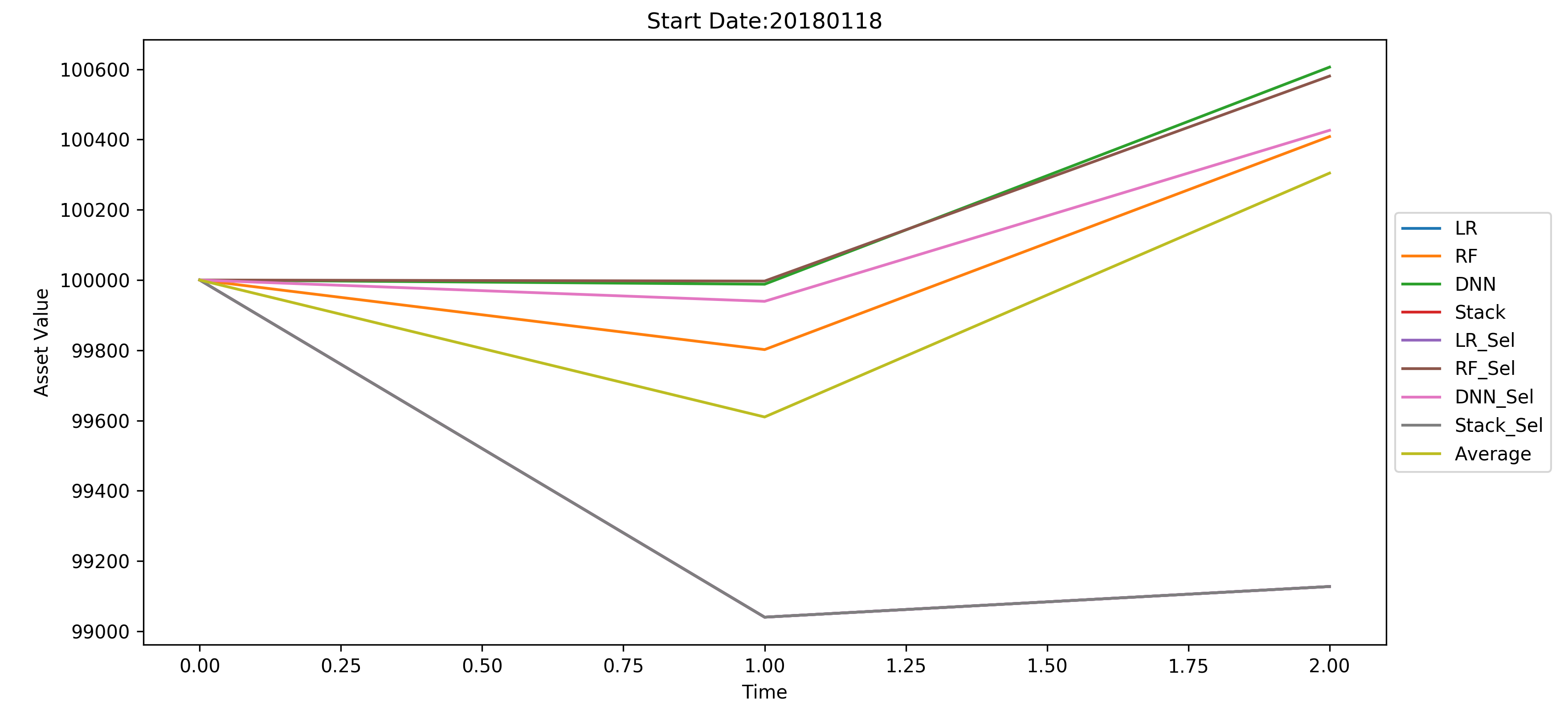}
	\caption{Portfolio Management starts from 2018.01.18}\label{fig 7}
\end{figure}

To this end, for each model, we use its output, which is a vector of scalers ranging from zero to one, to construct portfolio and trade it in historical market competing with the market average. \hyperref[fig 5]{Fig. 5.}, \hyperref[fig 6]{Fig. 6.}, and \hyperref[fig 7]{Fig. 7.} demonstrate the back test results showing that our stock selection strategy can construct profitable portfolios conquering the market average.

\vspace{0.5cm}

\section*{Acknowledgment}
We would like to say thanks to YuYan Shi, ZiYi Yang, MingDi Zheng, and Yuan Zeng for their incipient involvement of the project. We also thank Professor ZhiHong Huang of Sun Yat-sen University for his patient and generous help throughout the research.

\onecolumn
\appendix \label{Appendix}
\vspace{1.5cm}
\centering
{\tiny\renewcommand{\arraystretch}{.8}
\resizebox{!}{.35\paperheight}{
\begin{supertabular}[\textwidth]{c|c|c|c}
  \toprule
  AccountsPayablesTDays	&AccountsPayablesTRate	&AdminiExpenseRate	&ARTDays	\\
  AdminiExpenseRate	&ARTDays	&ARTRate	&ASSI	\\
  ARTRate	&ASSI	&BLEV	&BondsPayableToAsset	\\
  BLEV	&BondsPayableToAsset	&CashRateOfSales	&CashToCurrentLiability	\\
  CashRateOfSales	&CashToCurrentLiability	&CMRA	&CTOP	\\
  CMRA	&CTOP	&CTP5	&CurrentAssetsRatio	\\
  CTP5	&CurrentAssetsRatio	&CurrentAssetsTRate	&CurrentRatio	\\
  CurrentAssetsTRate	&CurrentRatio	&DAVOL10	&DAVOL20	\\
  DAVOL10	&DAVOL20	&DAVOL5	&DDNBT	\\
  DAVOL5	&DDNBT	&DDNCR	&DDNSR	\\
  DDNCR	&DDNSR	&DebtEquityRatio	&DebtsAssetRatio	\\
  DebtEquityRatio	&DebtsAssetRatio	&DHILO	&DilutedEPS	\\
  DHILO	&DilutedEPS	&DVRAT	&EBITToTOR	\\
  DVRAT	&EBITToTOR	&EGRO	&EMA10	\\
  EGRO	&EMA10	&EMA120	&EMA20	\\
  EMA120	&EMA20	&EMA5	&EMA60	\\
  EMA5	&EMA60	&EPS	&EquityFixedAssetRatio	\\
  EPS	&EquityFixedAssetRatio	&EquityToAsset	&EquityTRate	\\
  EquityToAsset	&EquityTRate	&ETOP	&ETP5	\\
  ETOP	&ETP5	&FinancialExpenseRate	&FinancingCashGrowRate	\\
  FinancialExpenseRate	&FinancingCashGrowRate	&FixAssetRatio	&FixedAssetsTRate	\\
  FixAssetRatio	&FixedAssetsTRate	&GrossIncomeRatio	&HBETA	\\
  GrossIncomeRatio	&HBETA	&HSIGMA	&IntangibleAssetRatio	\\
  HSIGMA	&IntangibleAssetRatio	&InventoryTDays	&InventoryTRate	\\
  InventoryTDays	&InventoryTRate	&InvestCashGrowRate	&LCAP	\\
  InvestCashGrowRate	&LCAP	&LFLO	&LongDebtToAsset	\\
  LFLO	&LongDebtToAsset	&LongDebtToWorkingCapital	&LongTermDebtToAsset	\\
  LongDebtToWorkingCapital	&LongTermDebtToAsset	&MA10	&MA120	\\
  MA10	&MA120	&MA20	&MA5	\\
  MA20	&MA5	&MA60	&MAWVAD	\\
  MA60	&MAWVAD	&MFI	&MLEV	\\
  MFI	&MLEV	&NetAssetGrowRate	&NetProfitGrowRate	\\
  NetAssetGrowRate	&NetProfitGrowRate	&NetProfitRatio	&NOCFToOperatingNI	\\
  NetProfitRatio	&NOCFToOperatingNI	&NonCurrentAssetsRatio	&NPParentCompanyGrowRate	\\
  NonCurrentAssetsRatio	&NPParentCompanyGrowRate	&NPToTOR	&OperatingExpenseRate	\\
  NPToTOR	&OperatingExpenseRate	&OperatingProfitGrowRate	&OperatingProfitRatio	\\
  OperatingProfitGrowRate	&OperatingProfitRatio	&OperatingProfitToTOR	&OperatingRevenueGrowRate	\\
  OperatingProfitToTOR	&OperatingRevenueGrowRate	&OperCashGrowRate	&OperCashInToCurrentLiability	\\
  OperCashGrowRate	&OperCashInToCurrentLiability	&PB	&PCF	\\
  PB	&PCF	&PE	&PS	\\
  PE	&PS	&PSY	&QuickRatio	\\
  PSY	&QuickRatio	&REVS10	&REVS20	\\
  REVS10	&REVS20	&REVS5	&ROA	\\
  REVS5	&ROA	&ROA5	&ROE	\\
  ROA5	&ROE	&ROE5	&RSI	\\
  ROE5	&RSI	&RSTR12	&RSTR24	\\
  RSTR12	&RSTR24	&SalesCostRatio	&SaleServiceCashToOR	\\
  SalesCostRatio	&SaleServiceCashToOR	&SUE	&TaxRatio	\\
  SUE	&TaxRatio	&TOBT	&TotalAssetGrowRate	\\
  TOBT	&TotalAssetGrowRate	&TotalAssetsTRate	&TotalProfitCostRatio	\\
  TotalAssetsTRate	&TotalProfitCostRatio	&TotalProfitGrowRate	&VOL10	\\
  TotalProfitGrowRate	&VOL10	&VOL120	&VOL20	\\
  VOL120	&VOL20	&VOL240	&VOL5	\\
  VOL240	&VOL5	&VOL60	&WVAD	\\
  VOL60	&WVAD	&REC	&DAREC	\\
  REC	&DAREC	&GREC	&FY12P	\\
  GREC	&FY12P	&DAREV	&GREV	\\
  DAREV	&GREV	&SFY12P	&DASREV	\\
  SFY12P	&DASREV	&GSREV	&FEARNG	\\
  GSREV	&FEARNG	&FSALESG	&TA2EV	\\
  FSALESG	&TA2EV	&CFO2EV	&ACCA	\\
  \bottomrule
\end{supertabular}}

\end{document}